\documentclass[12pt]{article}        

\usepackage{amsmath}
\usepackage{amssymb}

\usepackage{epsfig}

\usepackage{cite}

\usepackage{epic}
\usepackage{fancybox}

\usepackage{html}

\begin{document}

\begin{titlepage}

\begin{flushright}
        {\sf\large  Version of  \today{} \\
}\end{flushright}

\vspace{10mm}
\begin{center}
{\LARGE\bf
Practical Guide to Monte Carlo
}
\end{center}
 
\begin{center}
  {\bf S. Jadach}\\
   {\em Institute of Nuclear Physics,
        ul. Kawiory 26a, Krak\'ow, Poland}\\
   { and}\\
   {\em  DESY, Theory Group, Notkestrasse 85,  D-22603 Hamburg, Germany}\\
\end{center}

\vspace{10mm}
\begin{abstract}
I show how to construct Monte Carlo algorithms (programs),
prove that they are correct and document them.
Complicated algorithms are build using a handful of elementary methods.
This construction process is transparently illustrated using
graphical representation in which complicated graphs
consist of only several elementary building blocks.
In particular I discuss the equivalent algorithms,
that is different MC algorithms, 
with  different arrangements of the elementary building blocks,
which generate {\em the same} final probability distribution.
I also show how to transform a given MC algorithm into another equivalent one
and  discuss advantages of the various ``architectures''.
\end{abstract}

\begin{center}
  {\em To be submitted somewhere, sometime (or may be not)}
\end{center}

\end{titlepage}

\section{Introduction}

The aim of this report is to provide:
\begin{itemize}
\item
  Elementary description of elementary Monte Carlo (MC) methods for a graduate student
  in physics, who is supposed to learn them in a couple of days,
\item
  Methods of transparent documenting of the MC programs,
\item
  Reference in publications where there is no space for description
  of the elementary MC methodology.
\end{itemize}
In my opinion there is certain gap in the published literature on the MC methods.
The elementary MC methods like rejection according to weight,
branching (multichannel method) or mapping of variables are so simple 
and intuitive that it seems to be not worth to write anything on them.
On the other hand in the practical MC applications these methods
are often combined in such a complicated and baroque way that
sometimes one may wonder if the author is really controlling what he is
doing, especially if the documentation is incomplete/sparse and we lack
commonly accepted terminology graphical notation for describing MC algorithms.
There are also many mathematically oriented articles and textbooks
on the MC methods which in my opinion seem to have very little connection with
the practical every day work of someone constructing MC program.
The aim of this report is to fill at least partly this gap.
This report is extension of a section in 
ref.~\cite{jadach:1985}.
I would like also to recommend the classical report 
\cite{James:1980} of James on the elementary MC methods.

Section 1 describes elementary MC methods of the single and multiple level
rejection (reweighting), including detailed description of the weight book-keeping
and recipes for keeping correct normalisation for the total integrand 
and the differential
distributions (histograms).
Section 2 introduces branching (multi-channel) method and section 3 demonstrates
the simplest combinations of the rejection and branching.
Section 4 and 5 reviews more advanced aspects of combining rejection and branching,
in particular I show examples of ``equivalent'' algorithms, i.e.
different algorithm which provide the same distributions,
pointing out advantages of certain arrangements of the the rejection and branching.
Another common method of the variable mapping
is discussed in section 5, again in the context of various arrangement
of the rejection and branching methods.

\newpage
\section{Rejection, compensating weights}

We intend to generate randomly {\em events}, 
that is points $x=(x_1,x_2,x_3,...,x_n)$, according to a distribution
\begin{equation}
\rho(x_i) = { d^n\sigma \over d x^n }(x_i)
\end{equation}
within certain domain $\Omega$
and, simultaneously, we want to calculate (estimate) the integral
\begin{equation}
\sigma = \int\limits_\Omega \rho(x_i)\; dx^n.
\end{equation}
as precisely as possible.
In our notation, change of integration variables induces in $\rho$-density
a {\em Jacobian} factor%
\footnote{ 
  I assume that the reader is familiar only with the elementary
  calculus methods, and purposely avoid to call $\rho$ a measure.}
\begin{equation}
\rho(y_i) = \left| { \partial x \over \partial y } \right|\; \rho(x_i).
\end{equation}
The normalised to unity probability density is simply
\begin{equation}
d^n p = {1\over \sigma } d^n \sigma,\qquad
\int\limits_\Omega d^n p =
\int\limits_\Omega {d^n p \over dx^n}\; dx^n =1.
\end{equation}
%

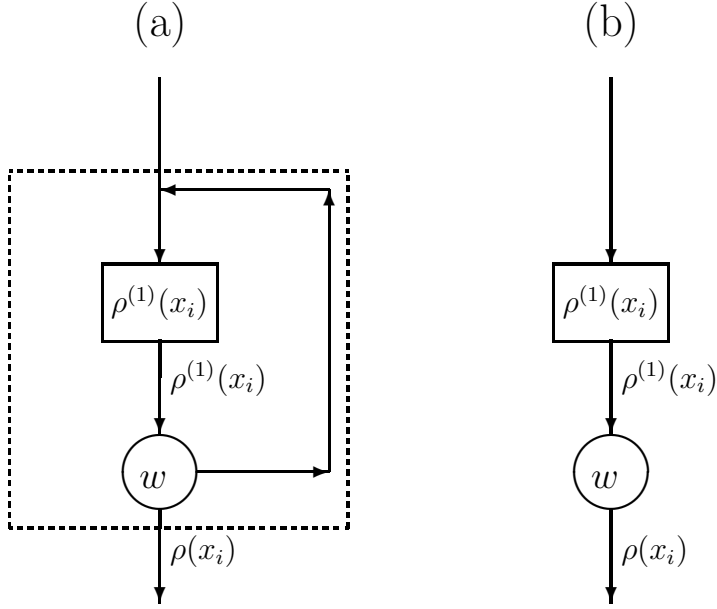
\begin{figure}[!ht]
\centering
\htmlimage{scale=1.4}
\setlength{\unitlength}{0.1mm}
\begin{picture}(1400,850)
%
%
\put( 400,800){\makebox(0,0)[t]{\hbox{\Large (a)}}}
\put( 400,850){\makebox(0,0)[t]{\begin{picture}( 600,850)
\thicklines
%
\put(300,700){\vector(0,-1){250}}
\put(300,450){\makebox(0,0)[t]{\framebox(150,100){$\rho^{(1)}(x_i)$}}}
\drawline(300, 350)( 300, 300)          
\put(300,300){\makebox(0,0)[t]{\begin{picture}( 600,200)
      \put(300, 200){\vector(0,-1) {75}}   
      \put(300,  75){\circle{100}}
      \put(300,  75){\makebox(0,0)[t]{\hbox{\large $w$ }}}
      \drawline(300, 25)( 300, 0)          
      \put(350, 75){\vector( 1, 0){175}} 
      \put(525, 75){\vector( 0, 1){375}} 
      \put(525,450){\vector(-1, 0){225}} 
\end{picture}}}
\put(300,100){\vector(0,-1){100}} 
\put(100,100){\dashbox{7}(450,475){ }}
\put(300,300){\makebox(0,0)[l]{\hbox{ $\rho^{(1)}(x_i)$ }}}
\put(300, 70){\makebox(0,0)[l]{\hbox{ $\rho(x_i)$ }}}
\thinlines
\end{picture}}}
%
%
%
\put(1000,800){\makebox(0,0)[t]{\hbox{\Large (b)}}}
\put(1000,850){\makebox(0,0)[t]{\begin{picture}( 600,850)
\thicklines
%
\put(300,700){\vector(0,-1){250}}
\put(300,450){\makebox(0,0)[t]{\framebox(150,100){$\rho^{(1)}(x_i)$}}}
\drawline(300, 350)( 300, 300)          
\put(300,300){\makebox(0,0)[t]{\begin{picture}( 600,200)
      \put(300, 200){\vector(0,-1) {75}}   
      \put(300,  75){\circle{100}}
      \put(300,  75){\makebox(0,0)[t]{\hbox{\large $w$ }}}
      \drawline(300, 25)( 300, 0)          
\end{picture}}}
\put(300,100){\vector(0,-1){100}} 
\put(300,300){\makebox(0,0)[l]{\hbox{ $\rho^{(1)}(x_i)$ }}}
\put(300, 70){\makebox(0,0)[l]{\hbox{ $\rho(x_i)$ }}}
\thinlines
\end{picture}}}
%
%
%
%
\end{picture} 
\caption{\sf\small
Single-line MC algorithm:
versions (a) constant-weight events
and (b) variable-weight events.
}
\label{fig:simple1}
\end{figure}

In fig. \ref{fig:simple1}(a) I show the simple single-line
algorithm with rejection according to a {\em weight}%
\footnote{ 
  I assume that the distribution $\rho$ may be quite singular, in particular
  it may include Dirac $\delta$ functions. 
  The weight I expect to be analytical almost everywhere.
  It may, however, include discontinuities.}
defined as a ratio of an {\em exact} distribution 
$\rho$ to {\em approximate} $\rho^{(1)}$
\begin{equation}
w(x) = { \rho(x_i) \over \rho^{(1)}(x_i)}
     = { d^n \sigma \over d^n \sigma^{(1)} }.
\end{equation}
We assume that we are able to generate randomly events 
according to $\rho^{(1)}$ and we know the numerical value of the integral
\begin{equation}
\sigma^{(1)} = \int\limits_\Omega \rho(x_i)^{(1)}\; dx^n.
\end{equation}
The box \fbox{$\rho^{(1)}(x_i)$} represents part of the algorithm
(part of computer code) which provides us events according to
$\rho^{(1)}$ and the value of $\sigma^{(1)}$.
The content of the box \fbox{$\rho^{(1)}(x_i)$} can be a complicated
algorithm (computer code)
and we treat it as a ``black box'', i.e. we may know
nothing about its content.
In particular it can be taken from ready-to-use library of the programs
generating standard distributions, or even a physical process
providing ``random events''.
The circle with the return line depicts {\em rejection method}.
For each event leaving \fbox{$\rho^{(1)}(x_i)$}
we calculate weight $w$ and we accept event (downward arrow) if
\begin{equation}
 r W < w,
\end{equation}
where W is a {\em maximum weight} and $r$ is a 
{\em uniform random number} $ 0<r<1.$
Otherwise event is rejected (return arrow in the graph).
It is easy to see that events exiting our algorithm are
generated according to density $\rho$.
Probability density of accepted events at the point $x_i$ is
equal to product of probability $d^n p^{(1)}$ of events produced in the
box $\rho^{(1)}$ times the probability 
$p_{accept}=w(x)/W$ of accepting an event
\begin{equation}
 d^n p(x) = {\cal N} d^n p^{(1)}\; p_{accept} = 
          {\cal N} { d^n \sigma^{(1)}(x) \over \sigma^{(1)} }\;
          { w(x) \over W},
\end{equation}
where ${\cal N}$ is normalisation factor.
Substituting the definition of the weight $w(x)=d^n\sigma/d^n\sigma^{(1)}$ and
imposing normalisation condition $\int d^n p(x) =1$ we find 
\begin{equation}
 d^n p(x) = { {\cal N} \over W}\; { d^n \sigma(x) \over \sigma^{(1)} },
 \qquad
 {\cal N} = { W \sigma^{(1)} \over \sigma}
\end{equation}
and as a result we obtain
\begin{equation}
 d^n p(x) = { d^n \sigma(x) \over \sigma }
\end{equation}
as desired.
The dashed box $\rho(x_i)$ can be used as part in a bigger algorithm
(box in a bigger graph) because it provides events generated
according to density $\rho(x_i)$.
The question is whether within the dashed box we are able to estimate
the integral $\sigma$.
In fact we can, and there are even two ways to do it.
In the first method we use the ratio on accepted events $N$ to
the total number $N^{(1)}$ of the events generated in the 
box \fbox{$\rho^{(1)}(x_i)$}.
The number of accepted events is, on the average, proportional to
probability of generating an event  event $d^n \sigma^{(1)}/\sigma^{(1)}$
times probability of accepting an event
\begin{equation}
\bar{w}={ w(x) \over W},\qquad 
\end{equation}
averaged all over the points $x_i$ in the entire 
integration (generation) domain $\Omega$
\begin{equation}
\label{naccept}
 N = N^{(1)} \int\limits_\Omega 
     {d^n \sigma^{(1)} \over \sigma^{(1)}}\; \bar{w}
   = N^{(1)} { \sigma \over \bar{\sigma}^{(1)} },\quad
     \bar{\sigma}^{(1)}= W\;\sigma^{(1)}.
\end{equation}
The above relation can be used to calculate the unknown integral $\sigma$ as follows
\begin{equation}
\label{norm-nacc}
 \sigma = \bar{\sigma}^{(1)}\;  { N \over N^{(1)} }.
\end{equation}
using known $\sigma^{(1)}$ and counting accepted events $N$.
Of course, the error of the above estimator of $\sigma$ is given
by the usual statistical
error from the binomial distribution.
In the second method we calculate the average weight where the averaging
is done over all accepted and rejected events
\begin{equation}
 <w> =  \int\limits_\Omega 
     {d^n \sigma^{(1)} \over \sigma^{(1)}}\; w(x)
     = { \sigma \over \sigma^{(1)} }.
\end{equation}
The above gives us second equivalent estimator of the unknown integral $\sigma$
in terms of the known $\sigma^{(1)}$ and the measured average weight
\begin{equation}
\label{norm-avewt}
  \sigma = \sigma^{(1)}\;  <w> = \bar{\sigma}^{(1)} \;  <\bar{w}>.
\end{equation}
Another often asked question is: how to calculate the
integral $\Delta\sigma$ over a {\em subdomain} $\Delta\Omega$,
which is for instance a single bin in a histogram?
The following formula can be easily derived
\begin{equation}
  \Delta\sigma = \sigma\; {\Delta N \over N}
               = \bar{\sigma}^{(1)} {\Delta N \over N^{(1)}}
\end{equation}
where $\Delta N$ is number of events falling into subdomain $\Delta\Omega$.
A particular case is the proper normalisation of the histogram.
Let us take the one dimensional distribution
\begin{equation}
  {d\sigma \over d z} = \int\limits_\Omega d^n\sigma\;  \delta(z - z(x_i))
\end{equation}
which we estimate/calculate by means of collecting generated
events in a histogram with $n_b$ equal bins within 
a $(z_{\min},z_{\max})$ range.
The relevant formula reads
\begin{equation}
  {d\sigma \over d z} 
  \simeq {\sigma \Delta N   \over  \Delta z N  }
  =  {  n_b \sigma \Delta N    \over    (z_{\max}-z_{\min}) N}
  =  {  n_b \bar{\sigma}^{(1)} \Delta N \over (z_{\max}-z_{\min}) N^{(1)} }
\end{equation}

In fig. \ref{fig:simple1}(b) I show the same algorithm
for variable-weight events.
In this case we do not reject events but we associate the weight
$w$ with each event.
For the total integral over entire $\Omega$ I may use the same
formula of eq. (\ref{norm-avewt}) as for the constant-weight algorithm.
In the case of the histogram we accumulate in each bin a sum of weights
$\sum\limits_{z\in bin} w$.
The properly normalised distribution is obtained as 
follows\footnote{Operationally this formula is identical for
  the case of variable-weight and constant-weight events
  and is therefore handy in practical calculations \cite{bhlumi2:1992}.
  The values of $\bar{\sigma}^{(1)}$ and $N^{(1)}$
  can be accumulated using a dedicated, one-bin-histogram.
  This arrangement facilitates dumping all MC results on
  the disk and restarting MC run at the later time.}
\begin{equation}
  {d\sigma \over d z} 
  = {  n_b \sigma^{(1)}        \over (z_{\max}-z_{\min}) N^{(1)} }
       \sum\limits_{z\in bin} w
  = {  n_b  \bar{\sigma}^{(1)}  \over (z_{\max}-z_{\min}) N^{(1)} }
       \sum\limits_{z\in bin} \bar{w}
\end{equation}
%

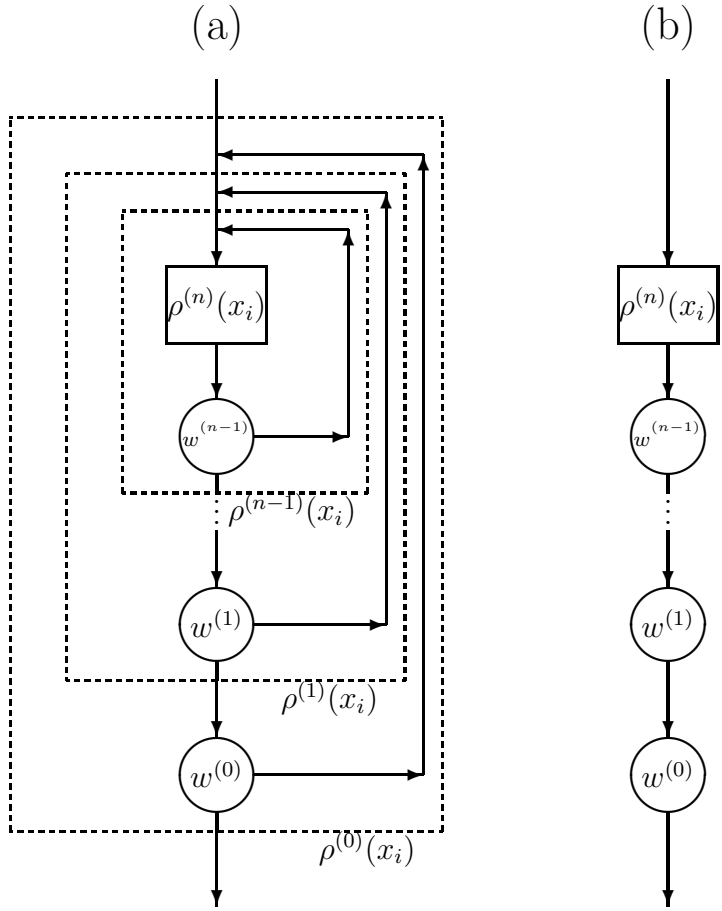
\begin{figure}[!ht]
\centering
\htmlimage{scale=1.4}
\setlength{\unitlength}{0.1mm}
\begin{picture}(1400,1200)
%
%
\put( 400,1200){\makebox(0,0)[t]{\hbox{\Large (a)}}}
\put( 400,1100){\makebox(0,0)[t]{\begin{picture}( 600,1100)
\thicklines
\put(175,550){\dashbox{7}(325,375){ }}
\put(400,550){\makebox(0,0)[t]{\hbox{ $\rho^{(n-1)}(x_i)$ }}}
\put(100,300){\dashbox{7}(450,675){ }}
\put(450,300){\makebox(0,0)[t]{\hbox{ $\rho^{(1)}(x_i)$ }}}
\put( 25,100){\dashbox{7}(575,950){ }}
\put(500,100){\makebox(0,0)[t]{\hbox{ $\rho^{(0)}(x_i)$ }}}
%
\put(300,1100){\vector(0,-1){250}}
\put(300,850){\makebox(0,0)[t]{\framebox(130,100){$\rho^{(n)}(x_i)$}}}
\put(300,750){\makebox(0,0)[t]{\begin{picture}( 600,200)
      \put(300, 200){\vector(0,-1) {75}}   
      \put(300,  75){\circle{100}}
      \put(300,  95){\makebox(0,0)[t]{\hbox{\scriptsize  $w^{^{(n-1)}}$}}}
      \drawline(300, 25)( 300, 0)          
      \put(350, 75){\vector( 1, 0){125}} 
      \put(475, 75){\vector( 0, 1){275}} 
      \put(475,350){\vector(-1, 0){175}} 
\end{picture}}}
\put(300,562){\makebox(0,0)[t]{\hbox{\vdots}}}
\put(300,500){\makebox(0,0)[t]{\begin{picture}( 600,200)
      \put(300, 200){\vector(0,-1) {75}}   
      \put(300,  75){\circle{100}}
      \put(300,  95){\makebox(0,0)[t]{\hbox{ $w^{(1)}$ }}}
      \drawline(300, 25)( 300, 0)          
      \put(350, 75){\vector( 1, 0){175}} 
      \put(525, 75){\vector( 0, 1){575}} 
      \put(525,650){\vector(-1, 0){225}} 
\end{picture}}}
\put(300,300){\makebox(0,0)[t]{\begin{picture}( 600,200)
      \put(300, 200){\vector(0,-1) {75}}   
      \put(300,  75){\circle{100}}
      \put(300,  95){\makebox(0,0)[t]{\hbox{ $w^{(0)}$ }}}
      \drawline(300, 25)( 300, 0)          
      \put(350, 75){\vector( 1, 0){225}} 
      \put(575, 75){\vector( 0, 1){825}} 
      \put(575,900){\vector(-1, 0){275}} 
\end{picture}}}
\put(300,100){\vector(0,-1){100}} 
\thinlines
\end{picture}}}
%
%
%
\put(1000,1200){\makebox(0,0)[t]{\hbox{\Large (b)}}}
\put(1000,1100){\makebox(0,0)[t]{\begin{picture}( 600,1100)
\thicklines
%
\put(300,1100){\vector(0,-1){250}}
\put(300,850){\makebox(0,0)[t]{\framebox(130,100){$\rho^{(n)}(x_i)$}}}
\put(300,750){\makebox(0,0)[t]{\begin{picture}( 600,200)
      \put(300, 200){\vector(0,-1) {75}}   
      \put(300,  75){\circle{100}}
      \put(300,  95){\makebox(0,0)[t]{\hbox{\scriptsize  $w^{^{(n-1)}}$}}}
      \drawline(300, 25)( 300, 0)          
\end{picture}}}
\put(300,562){\makebox(0,0)[t]{\hbox{\vdots}}}
\put(300,500){\makebox(0,0)[t]{\begin{picture}( 600,200)
      \put(300, 200){\vector(0,-1) {75}}   
      \put(300,  75){\circle{100}}
      \put(300,  95){\makebox(0,0)[t]{\hbox{ $w^{(1)}$ }}}
      \drawline(300, 25)( 300, 0)          
\end{picture}}}
\put(300,300){\makebox(0,0)[t]{\begin{picture}( 600,200)
      \put(300, 200){\vector(0,-1) {75}}   
      \put(300,  75){\circle{100}}
      \put(300,  95){\makebox(0,0)[t]{\hbox{ $w^{(0)}$ }}}
      \drawline(300, 25)( 300, 0)          
\end{picture}}}
\put(300,100){\vector(0,-1){100}} 
\thinlines
\end{picture}}}
%
%
%
%
\end{picture} 
\caption{\sf\small
Single-line MC algorithm:
versions (a) with constant-weight events, nested rejection loops,
and (b) with variable-weight events.
}
\label{fig:single-line}
\end{figure}

In fig. \ref{fig:single-line}(a) I show the simple single-line
algorithm with several nested rejection loops.
The meaning of the graph is rather obvious.
The original distribution $\rho_0$ goes through $n$-step
simplification procedure
\begin{equation}
\rho^{(0)} \to \rho^{(1)} \to \rho^{(2)} \dots \to \rho^{(n)}
\end{equation}
and the compensation weights
\begin{equation}
w^{(k)} = { \rho^{(k)} \over \rho^{(k-1)}}
\end{equation}
are used for rejections ``locally'' in a standard way:
each weight $w^{(k)}$ is compared with $r W^{(k)}$
where $0<r<1$ is uniform random number,
and if  $w^{(k)}< r W^{(k)}$ the event is accepted
(down-ward arrow), otherwise rejected (return loop).
The average weights $<w^{(k)}>$ are calculated for each rejection loop.
The most inward box \fbox{$\rho^{(n)}(x_i)$} represents
generation algorithm of the points $x_i$
according to maximally simplified (crude) distribution $\rho^{(n)}$
for which we know the integral 
$\sigma^{(n)}=\int \rho^{(n)}$ 
analytically.
The integral of the original distribution $\rho$ is obtained
from the crude integral and the average weights
\begin{equation}
\sigma^{(0)} = \int \rho(x_i)\; dx^n
   = \sigma^{(n)}\; \prod\limits_{i=1}^n <w^{(i-1)}>
   = \bar{\sigma}^{(n)}\; \prod\limits_{i=1}^n <\bar{w}^{(i-1)}>
\end{equation}
The above is completely standard and can be found in
ref.  \cite{jadach:1985}.
Note also that all $n$ rejection loops may be combined into
single rejection loop with the weight being product
of all weights along the line
\begin{equation}
w   = \prod\limits_{i=1}^n w^{(i-1)}.
\end{equation}
Usually, the version with nested loops is more efficient
and the corresponding program is more modular.
The weights for the internal loops are related to more
technical aspects of the MC algorithm (Jacobians)
and do not evolve quickly (during the development of the program)
while external weights correspond to physics model and may change more frequently.
It is therefore profitable to keep in practice several
levels of the weights.

Finally, we may decide to perform calculation for
weighted events.
In fig.~\ref{fig:single-line}(b) I show the version of the
simple single-line MC algorithm with variable-weight events.
The event at the exit of the graph gets associated weight $w$
which is the product of all weights along the line.

\section{Branching}
\label{sect:branching}

\begin{figure}[!ht]
\centering
\htmlimage{scale=1.4}
\setlength{\unitlength}{0.1mm}
\begin{picture}(1200,800)
%
%
\put(600,800){\makebox(0,0)[t]{\begin{picture}( 600,800)
%
\put(  0,150){\dashbox{10}(625,600){ }}
\put(500,150){\makebox(0,0)[t]{\hbox{ $\rho(x_i)$ }}}
\thicklines
\put(300,800){\makebox(0,0)[t]{\begin{picture}( 600,200)
      \put(300,200){\vector(0,-1){175}}
      \put(300, 00){\circle*{50}}
      \put(280, 15){\makebox(0,0)[r]{\hbox{ $P_k$ }}}
      \put(330, 15){\makebox(0,0)[l]{\hbox{\scriptsize $k=1,...,n$ }}}
\end{picture}}}
\put(300,600){\makebox(0,0)[t]{\begin{picture}( 600,100)
      \drawline(300,100)(100,0)
      \drawline(300,100)(300,0)
      \drawline(300,100)(500,0)
\end{picture}}}
\put(300,500){\makebox(0,0)[t]{\begin{picture}( 600,100)
      \put(100,100){\makebox(0,0)[t]{\framebox(130,100){$\rho_1(x_i)$}}}
      \put(300,100){\makebox(0,0)[t]{\framebox(130,100){$\rho_2(x_i)$}}}
      \put(500,100){\makebox(0,0)[t]{\framebox(130,100){$\rho_n(x_i)$}}}
      \put(400, 50){\makebox(0,0)[t]{\hbox{ \dots }}}
\end{picture}}}
\put(300,400){\makebox(0,0)[t]{\begin{picture}( 600,50)
      \put(100,50){\vector(0,-1){50}}
      \put(300,50){\vector(0,-1){50}}
      \put(500,50){\vector(0,-1){50}}
\end{picture}}}
\put(300,350){\makebox(0,0)[t]{\begin{picture}( 600,100)
      \drawline(100,100)(300,0)
      \drawline(300,100)(300,0)
      \drawline(500,100)(300,0)
\end{picture}}}
\put(300,250){\vector(0,-1){250}} 
\thinlines
\end{picture}}}
%
%
%
\end{picture} 
\caption{\sf\small
Pure branching. 
}
\label{fig:branching}
\end{figure}
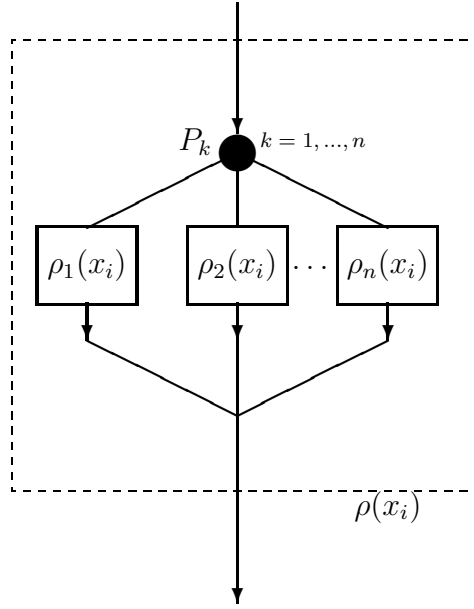

In fig. \ref{fig:branching} I show the general 
MC algorithm with branching into $n$ branches.
This kind of algorithm is used when the distribution
to be generated $\rho$ can be split into sum of
several distinct (positive) subdistributions
\begin{equation}
\rho(x_i) = \sum_{i=1}^n \rho_i
\end{equation}
and we are able, one way or another, to generate
each $\rho_i$ separately.
Usually each $\rho_i$ contains single peak
or one class of peaks in the distribution $\rho$.
In  the beginning of the algorithm (black disc) we pick randomly
one branch (subdistribution) according to probability
\begin{equation}
P_k = { \int \rho_k \over \sum\limits_i \int \rho_i }
    = {\sigma_k \over \sigma},
\end{equation}
i.e. we have to know {\em in advance} the integrals $\sigma_i= \int \rho_i$
analytically or numerically.
Typically, in each branch one uses different integration
variables $x_i$ to parameterise the integral.
The particular choice of variables will be adjusted
to leading ``singularities'' in the branch distribution.

Let us give a formal proof of the correctness of the branching method.
\begin{equation}
  d^n p(x) = \sum\limits_k P_k d^n p_k(x)
           = \sum\limits_k {\sigma_k \over \sigma}\;
                           {d^n \sigma_k(x)  \over \sigma_k}
           = {1\over \sigma} \sum\limits_k d^n \sigma_k(x)
           = {d^n \sigma(x) \over \sigma} 
\end{equation}

Finally, note that at the exit of the branched MC algorithm 
(exit of the graph in fig. \ref{fig:branching}),
we may be forced for various reason (saving computer memory), trash
all information on the origin of an event, consequently, we may be not able to
use any information specific to the branch from which
the event has came. 
This is rather important practical aspect to be kept in mind.

\section{Branching and internal rejection}

\begin{figure}[!ht]
\centering
\htmlimage{scale=1.4}
\setlength{\unitlength}{0.1mm}
\begin{picture}(1400,1100)
%
%
\put(300,1100){\makebox(0,0)[t]{\hbox{\Large (a)}}}
\put(300,1000){\makebox(0,0)[t]{\begin{picture}( 600,1100)
%
\put( 15,200){\dashbox{10}(600,850){ }}
\put(500,200){\makebox(0,0)[t]{\hbox{ $\rho(x_i)$ }}}
\thicklines
\put(300,1100){\makebox(0,0)[t]{\begin{picture}( 600,200)
      \put(300,200){\vector(0,-1){175}}
      \put(300, 00){\circle*{50}}
      \put(280, 15){\makebox(0,0)[r]{\hbox{ $P_k$ }}}
      \put(330, 15){\makebox(0,0)[l]{\hbox{\scriptsize $k=1,...,n$ }}}
\end{picture}}}
\put(300,900){\makebox(0,0)[t]{\begin{picture}( 600,100)
      \drawline(300,100)(100,0)
      \drawline(300,100)(300,0)
      \drawline(300,100)(500,0)
\end{picture}}}
\put(300,800){\makebox(0,0)[t]{\begin{picture}( 600,150)
      \put(100,150){\vector(0,-1){150}}
      \put(300,150){\vector(0,-1){150}}
      \put(500,150){\vector(0,-1){150}}
\end{picture}}}
\put(300,650){\makebox(0,0)[t]{\begin{picture}( 600,100)
      \put(100,100){\makebox(0,0)[t]{\framebox(130,100){$\rho^{(1)}_1(x_i)$}}}
      \put(300,100){\makebox(0,0)[t]{\framebox(130,100){$\rho^{(1)}_2(x_i)$}}}
      \put(500,100){\makebox(0,0)[t]{\framebox(130,100){$\rho^{(1)}_n(x_i)$}}}
      \put(400, 50){\makebox(0,0)[t]{\hbox{ \dots }}}
\end{picture}}}
\put(300,550){\makebox(0,0)[t]{\begin{picture}( 600,150)
      \put(100, 50){\makebox(0,0)[t]{\hbox{ $w_1$ }}}
      \put(100, 50){\circle{100}}
      \put(150, 50){\vector( 1, 0){ 50}}   
      \put(200, 50){\vector( 0, 1){275}}   
      \put(200,325){\vector(-1, 0){100}}   
      \put(300, 50){\makebox(0,0)[t]{\hbox{ $w_2$ }}}
      \put(300, 50){\circle{100}}
      \put(350, 50){\vector( 1, 0){ 50}}   
      \put(400, 50){\vector( 0, 1){275}}   
      \put(400,325){\vector(-1, 0){100}}   
      \put(500, 50){\makebox(0,0)[t]{\hbox{ $w_n$ }}}
      \put(500, 50){\circle{100}}
      \put(550, 50){\vector( 1, 0){ 50}}   
      \put(600, 50){\vector( 0, 1){275}}   
      \put(600,325){\vector(-1, 0){100}}   
      \put(100,150){\vector( 0,-1){50}}
      \put(300,150){\vector( 0,-1){50}}
      \put(500,150){\vector( 0,-1){50}}
\end{picture}}}
\put(300,400){\makebox(0,0)[t]{\begin{picture}( 600,50)
      \put(100,50){\vector(0,-1){50}}
      \put(300,50){\vector(0,-1){50}}
      \put(500,50){\vector(0,-1){50}}
\end{picture}}}
\put(300,350){\makebox(0,0)[t]{\begin{picture}( 600,100)
      \drawline(100,100)(300,0)
      \drawline(300,100)(300,0)
      \drawline(500,100)(300,0)
\end{picture}}}
\put(300,250){\vector(0,-1){150}}
\thinlines
\end{picture}}}
%
%
\put(1000,1100){\makebox(0,0)[t]{\hbox{\Large (b)}}}
\put(1000,1000){\makebox(0,0)[t]{\begin{picture}( 600,1100)
%
\put( 15,200){\dashbox{10}(600,850){ }}
\put(500,200){\makebox(0,0)[t]{\hbox{ $\rho(x_i)$ }}}
\thicklines
\put(300,1100){\makebox(0,0)[t]{\begin{picture}( 600,200)
      \put(300,200){\vector(0,-1){175}}
      \put(300, 00){\circle*{50}}
      \put(280, 15){\makebox(0,0)[r]{\hbox{ $\bar{P}_k^{(1)}$ }}}
      \put(330, 15){\makebox(0,0)[l]{\hbox{\scriptsize $k=1,...,n$ }}}
\end{picture}}}
\put(300,900){\makebox(0,0)[t]{\begin{picture}( 600,100)
      \drawline(300,100)(100,0)
      \drawline(300,100)(300,0)
      \drawline(300,100)(500,0)
\end{picture}}}
\put(300,800){\makebox(0,0)[t]{\begin{picture}( 600,100)
      \put(100,100){\makebox(0,0)[t]{\framebox(130,100){$\rho^{(1)}_1(x_i)$}}}
      \put(300,100){\makebox(0,0)[t]{\framebox(130,100){$\rho^{(1)}_2(x_i)$}}}
      \put(500,100){\makebox(0,0)[t]{\framebox(130,100){$\rho^{(1)}_n(x_i)$}}}
      \put(400, 50){\makebox(0,0)[t]{\hbox{ \dots }}}
\end{picture}}}
\put(300,700){\makebox(0,0)[t]{\begin{picture}( 600,300)
      \put(100, 50){\makebox(0,0)[t]{\hbox{ $w_1$ }}}
      \put(100, 50){\circle{100}}
      \put(150, 50){\vector(1,0){450}}
      \put(300,125){\makebox(0,0)[t]{\hbox{ $w_2$ }}}
      \put(300,125){\circle{100}}
      \put(350,120){\vector(1,0){250}}
      \put(500,200){\makebox(0,0)[t]{\hbox{ $w_n$ }}}
      \put(500,200){\circle{100}}
      \put(550,200){\vector(1,0){ 50}}
      \put(600, 50){\vector( 0, 1){550}}
      \put(600,600){\vector(-1, 0){300}}
      \put(100,300){\vector( 0,-1){200}}
      \put(300,300){\vector( 0,-1){125}}
      \put(500,300){\vector( 0,-1){ 50}}
      \put(300, 75){\vector( 0,-1){ 75}}
      \put(500,150){\vector( 0,-1){150}}
\end{picture}}}
\put(300,400){\makebox(0,0)[t]{\begin{picture}( 600,50)
      \put(100,50){\vector(0,-1){50}}
      \put(300,50){\vector(0,-1){50}}
      \put(500,50){\vector(0,-1){50}}
\end{picture}}}
\put(300,350){\makebox(0,0)[t]{\begin{picture}( 600,100)
      \drawline(100,100)(300,0)
      \drawline(300,100)(300,0)
      \drawline(500,100)(300,0)
\end{picture}}}
\put(300,250){\vector(0,-1){150}}
\thinlines
\end{picture}}}
%
%
\end{picture} 
\caption{\sf\small
Branching and weights. 
Two equivalent generation methods
with individual compensation weight for each branch.
The case 
(a) with local return loop for each branch, 
(b) with common return point before branching.
See text for more explanations how to to transform algorithm (a) into (b).
}
\label{fig:bran-wt}
\end{figure}
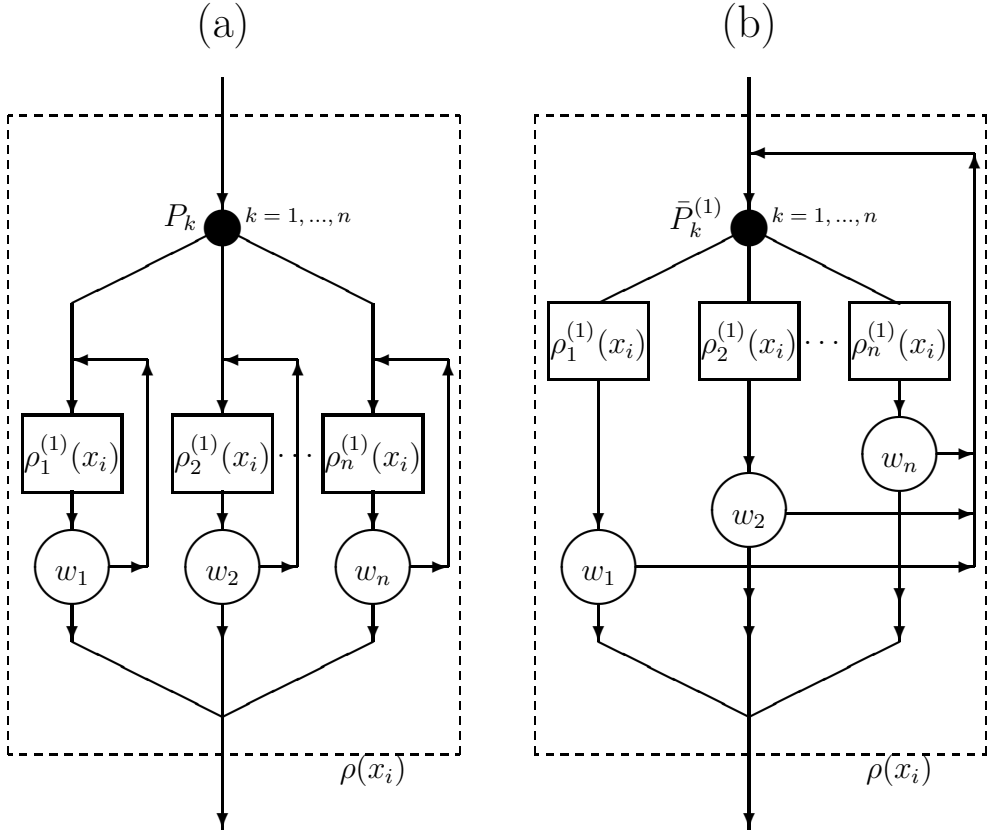

In fig. \ref{fig:bran-wt}(a) I show the simplest combination of the branching
algorithm with the standard rejection method.
This type of the algorithm is potentially very efficient but
is used not so often in the  MC event generators
because it requires that we know in advance branching probabilities $P_k$.
In most of situations we do not know them analytically.
In principle, they can be calculated numerically using
$\sigma_k=\bar{\sigma}^{(1)}_k <\bar{w}_k>$
but this is not very handy because it requires two MC runs --
first run which determines average weights $<\bar{w}_k>$
an second run with $P_k$ calculated from $<\bar{w}_k>$ from the first run.

The solution to the above problem is the variant of the
algorithm presented in fig. \ref{fig:bran-wt}(b)
where
\begin{equation}
  \bar{P}_k = {\bar{\sigma}^{(1)}_k \over \bar{\sigma}^{(1)} }
            = { \int \rho^{(1)}_k \over \sum\limits_l \int \rho^{(1)}_l }
\end{equation}
and all rejection returns are done to a point before the branching point.
Let us check correctness of the algorithm in fig. \ref{fig:bran-wt}(b).
The probability density $d^n p(x_i)$
at the point $x_i$ at the exit of the algorithm (graph)
is proportional to product of probability of getting
event in the box \fbox{$\rho^{(1)}_k(x_i)$} equal
$d^n p^{(1)}_k(x) = d^n \sigma^{(1)}_k(x) /\sigma_k^{(1)}$
times probability of accepting event being 
$\bar{w}_k(x)= d^n \sigma_k(x)/d^n \bar{\sigma}^{(1)}_k(x)$,
all that averaged over branches with probabilities $\bar{P}_k$.
The same statement is expressed mathematically as follows:
\begin{equation}
\begin{split}
d^n p(x_i) 
&= {\cal N} \sum\limits_k 
   \bar{P}_k\; d^n p^{(1)}_k(x)\; \bar{w}_k(x)
\\&
 = {\cal N} \sum\limits_k 
   {\bar{\sigma}_k^{(1)}  \over \bar{\sigma}^{(1)}  }\;
   {d^n \sigma^{(1)}_k(x) \over  \sigma_k^{(1)} }\;
   {d^n \sigma_k(x) \over  d^n \bar{\sigma}^{(1)}_k(x) }
 = {\cal N} { d^n\sigma(x) \over \bar{\sigma}^{(1)} }.
\end{split}
\end{equation}
Normalisation
${\cal N}= \bar{\sigma}^{(1)}/\sigma$
is determined from the condition 
$\int_\Omega d^n p(x_i) =1$.
Finally we obtain $d^n p(x_i) = d^n\sigma(x_i) / \sigma$, as expected.

The total integral is a sum over integrals from all branches
$\sigma=\sum_k \sigma_k$ where for each branch we may use
the formula $\sigma_k = \sigma^{(1)}_k <w_k>$ (see eq.~(\ref{norm-avewt})).
Slightly rearranged formula
\begin{equation}
\sigma = \sum_k \sigma^{(1)}_k <w_k>
       = \bar{\sigma}^{(1)} \sum_k \bar{P}_k <\bar{w}_k>
       = \bar{\sigma}^{(1)} <\bar{w}>,
\end{equation}
where in $<\bar{w}>$ we average also over branches,
is a straightforward  generalisation of eq.~(\ref{norm-avewt}).
We can also generalised formula of eq.~(\ref{norm-nacc}) for the total
integral based on the number of accepted events
\begin{equation}
\label{norm-bran-nacc}
\sigma = \bar{\sigma}^{(1)} { N \over N^{(1)} }
       = \bar{\sigma}^{(1)} { \sum_k N_k \over \sum_k N^{(1)}_k }
\end{equation}
{\em Proof}:
number $N$ of events accepted in all branches is
\begin{equation}
N =  \sum_k N_k = \sum_k N^{(1)}_k { \sigma_k \over \bar{\sigma}^{(1)}_k }
\end{equation}
where $N^{(1)}_k$ is total number of events in a given branch
(that is before rejection), see also eq.~(\ref{naccept}).
Inserting $N^{(1)}_k= N^{(1)} \bar{P}_k$ we get
$N = \sigma / \bar{\sigma}^{(1)}$ and therefore 
eq.~(\ref{norm-bran-nacc}).

Summarising, we see that the two algorithms in fig.~\ref{fig:bran-wt}
are equivalent, ie. they provide the same distribution
of events and the same total integral.
The algorithm (a) is probably slightly more efficient but also
more difficult to realize because it requires precise knowledge of 
the branching probabilities $P_k$.
The algorithm (b) is usually less efficient but 
the branching probabilities $\bar{P}^{(1)}_k$ are easier to evaluate
because they correspond to {\em simplified} distributions $\rho^{(1)}_k$.

Let us now consider the case of variable-weight events
for which all return loops in fig.~\ref{fig:bran-wt} are removed
and the two cases (a) and (b) are identical.
The event at the exit of the algorithm carries the weight
from one of the branches!
For the calculation of the total integral
we may use the same formulas of eqs.~(\ref{norm-bran-nacc}) and
(\ref{norm-avewt}) as for the constant-weight method.
Let us check whether we may proceed as usual for the calculation
of the integrals in the subdomain $\Delta\Omega$ being
single bin in any kind of the differential (or multi-differential)
distribution.
Let us generate long series of $N$ weighted events and
accumulate sum of the weights which fall into $\Delta\Omega$.
Of course, in the sum of accumulated weights
we have contributions from all branches
\begin{equation}
\sum\limits_{x_i \in \Delta\Omega} \bar{w}(x_i)
= \sum\limits_k \int\limits_{\Delta\Omega} d^n N_k\; \bar{w}_k
= \sum\limits_k N \bar{P}^{(1)}_k 
          \int\limits_{\Delta\Omega} d^n p^{(1)}_k\; \bar{w}_k
\end{equation}
Substituting definitions for $\bar{P}^{(1)}_k$ and of $d^n p^{(1)}_k$
we get
\begin{equation}
\begin{split}
\sum\limits_{x_i \in \Delta\Omega} \bar{w}(x_i)
&= N \sum\limits_k 
      { \bar{\sigma}^{(1)}_k        \over \bar{\sigma}^{(1)} }\;
  \int\limits_{\Delta\Omega}
      { d^n \bar{\sigma}^{(1)}_k(x) \over \bar{\sigma}^{(1)}_k }\;
      { d^n \sigma_k(x)       \over  d^n \bar{\sigma}^{(1)}_k(x) }
\\
&= {N \over \bar{\sigma}^{(1)} }\;
  \sum\limits_k \int\limits_{\Delta\Omega} d^n \sigma_k(x) 
= {N \over \bar{\sigma}^{(1)} }\; \Delta\sigma
\end{split}
\end{equation}
Reverting the above formula we get an estimate of the integrated
or (multi-) differential distribution in terms of sum of the weights
\begin{equation}
\Delta\sigma \equiv \int\limits_{\Delta\Omega} d^n \sigma_k
= { \bar{\sigma}^{(1)} \over N } 
  \sum\limits_{x_i \in \Delta\Omega} \bar{w}(x_i)
= { \sigma^{(1)} \over N } 
  \sum\limits_{x_i \in \Delta\Omega} w(x_i)
\end{equation}

Let us finally discuss the role of the maximum weight $W_k$ and
the apparently unnecessary complication of keeping two kinds
of crude distributions $\sigma^{(1)}$ and $\bar{\sigma}^{(1)}$.
For variable-weight events without branching, $W$ is merely
a scale factor which cancels out completely among $<\bar{w}>$
and $\bar{\sigma}^{(1)}$ in the overall normalisation.
Its only role is to keep weights in certain preferred range,
for example it is often preferred to have weights of order 1.
In the case of the variable-weights with branching the relative values
of $W_k$ start to play certain role.
Although, for infinite number of events, final results (distributions
and integrals) do not depend on $W_k$ the efficiency (convergence)
of the calculation depends on the relative ratios of $W_k$ \cite{Kleiss:1994qy}.
The maximum weights $W_k$ are more important/useful 
for constant-weight algorithm.
They are chosen in such a way that $\bar{w}<1$.
The rejection method does not work if this condition is not fulfilled.
In most cases we do not know analytically the maximum weight
for a given approximate $\rho^{(1)}$ and the maximum weight $W$ is adjusted
empirically.
Of course, the same adjustments can be done by scaling
(multiplying by a constant)
the entire $\rho^{(1)}$ but the long-standing tradition
tells us to keep $\rho^{(1)}$ unchanged and rather
introduce an explicit adjustment factor $W$.
In the case of the constant-weight algorithm the values of
$W_k$ determine the efficiency (rejection rate) in each branch.
Let us stress again that it is always possible to enforce $W_k=1$
and the presence of $W_k$ is in fact pure conventional.

\section{Branching and external rejection}

\begin{figure}[!ht]
\centering
\htmlimage{scale=1.4}
\setlength{\unitlength}{0.1mm}
\begin{picture}(1400,1100)
%
%
\put(300,1100){\makebox(0,0)[t]{\hbox{\Large (a)}}}
\put(300,1000){\makebox(0,0)[t]{\begin{picture}( 600,1100)
%
\put( 15,150){\dashbox{10}(600,900){ }}
\put(500,150){\makebox(0,0)[t]{\hbox{ $\rho(x_i)$ }}}
\thicklines
\put(300,1100){\makebox(0,0)[t]{\begin{picture}( 600,200)
      \put(300,200){\vector(0,-1){175}}
      \put(300, 00){\circle*{50}}
      \put(280, 15){\makebox(0,0)[r]{\hbox{ $\bar{P}_k^{(1)}$ }}}
      \put(330, 15){\makebox(0,0)[l]{\hbox{\scriptsize $k=1,...,n$ }}}
\end{picture}}}
\put(300,900){\makebox(0,0)[t]{\begin{picture}( 600,100)
      \drawline(300,100)(100,0)
      \drawline(300,100)(300,0)
      \drawline(300,100)(500,0)
\end{picture}}}
\put(300,800){\makebox(0,0)[t]{\begin{picture}( 600,100)
      \put(100,100){\makebox(0,0)[t]{\framebox(130,100){$\rho^{(1)}_1(x_i)$}}}
      \put(300,100){\makebox(0,0)[t]{\framebox(130,100){$\rho^{(1)}_2(x_i)$}}}
      \put(500,100){\makebox(0,0)[t]{\framebox(130,100){$\rho^{(1)}_n(x_i)$}}}
      \put(400, 50){\makebox(0,0)[t]{\hbox{ \dots }}}
\end{picture}}}
\put(300,700){\makebox(0,0)[t]{\begin{picture}( 600,300)
      \put(100, 50){\makebox(0,0)[t]{\hbox{ $w_1$ }}}
      \put(100, 50){\circle{100}}
      \put(150, 50){\vector(1,0){450}}
      \put(300,125){\makebox(0,0)[t]{\hbox{ $w_2$ }}}
      \put(300,125){\circle{100}}
      \put(350,120){\vector(1,0){250}}
      \put(500,200){\makebox(0,0)[t]{\hbox{ $w_n$ }}}
      \put(500,200){\circle{100}}
      \put(550,200){\vector(1,0){ 50}}
      \put(600, 50){\vector( 0, 1){550}}
      \put(600,600){\vector(-1, 0){300}}
      \put(100,300){\vector( 0,-1){200}}
      \put(300,300){\vector( 0,-1){125}}
      \put(500,300){\vector( 0,-1){ 50}}
      \drawline(300, 75)(300,0)
      \drawline(500,150)(500,0)
\end{picture}}}
\put(300,400){\makebox(0,0)[t]{\begin{picture}( 600,100)
      \put(100,100){\vector(0,-1){100}}
      \put(300,100){\vector(0,-1){100}}
      \put(500,100){\vector(0,-1){100}}
      \put(110, 60){\makebox(0,0)[l]{\hbox{\footnotesize $\rho_1(x_i)$ }}}
      \put(310, 60){\makebox(0,0)[l]{\hbox{\footnotesize $\rho_2(x_i)$ }}}
      \put(510, 60){\makebox(0,0)[l]{\hbox{\footnotesize $\rho_n(x_i)$ }}}
\end{picture}}}
\put(300,300){\makebox(0,0)[t]{\begin{picture}( 600,100)
      \drawline(100,100)(300,0)
      \drawline(300,100)(300,0)
      \drawline(500,100)(300,0)
\end{picture}}}
\put(300,250){\vector(0,-1){150}}
\thinlines
\end{picture}}}
%
%
\put(1000,900){\makebox(0,0)[t]{\hbox{\Large (b)}}}
\put(1000,800){\makebox(0,0)[t]{\begin{picture}( 600,800)
%
\put(  0,50){\dashbox{10}(625,700){ }}
\put(500,50){\makebox(0,0)[t]{\hbox{ $\rho(x_i)$ }}}
\put( 15,230){\dashbox{7}(570,410){ }}
\put(500,230){\makebox(0,0)[t]{\hbox{ $\rho^{(1)}(x_i)$ }}}
\thicklines
\put(300,800){\makebox(0,0)[t]{\begin{picture}( 600,200)
      \put(300,200){\vector(0,-1){175}}
      \put(300, 00){\circle*{50}}
      \put(280, 15){\makebox(0,0)[r]{\hbox{ $\bar{P}_k^{(1)}$ }}}
      \put(330, 15){\makebox(0,0)[l]{\hbox{\scriptsize $k=1,...,n$ }}}
\end{picture}}}
\put(300,600){\makebox(0,0)[t]{\begin{picture}( 600,100)
      \drawline(300,100)(100,0)
      \drawline(300,100)(300,0)
      \drawline(300,100)(500,0)
\end{picture}}}
\put(300,500){\makebox(0,0)[t]{\begin{picture}( 600,100)
      \put(100,100){\makebox(0,0)[t]{\framebox(130,100){$\rho^{(1)}_1(x_i)$}}}
      \put(300,100){\makebox(0,0)[t]{\framebox(130,100){$\rho^{(1)}_2(x_i)$}}}
      \put(500,100){\makebox(0,0)[t]{\framebox(130,100){$\rho^{(1)}_n(x_i)$}}}
      \put(400, 50){\makebox(0,0)[t]{\hbox{ \dots }}}
\end{picture}}}
\put(300,400){\makebox(0,0)[t]{\begin{picture}( 600,50)
      \put(100,50){\vector(0,-1){50}}
      \put(300,50){\vector(0,-1){50}}
      \put(500,50){\vector(0,-1){50}}
\end{picture}}}
\put(300,350){\makebox(0,0)[t]{\begin{picture}( 600,100)
      \drawline(100,100)(300,0)
      \drawline(300,100)(300,0)
      \drawline(500,100)(300,0)
\end{picture}}}
\put(300,250){\makebox(0,0)[t]{\begin{picture}( 600,250)
      \put(300,250){\vector(0,-1){50}}
      \put(300,150){\circle{100}}
      \put(300,150){\makebox(0,0)[t]{\hbox{ $w$ }}}
      \put(300,100){\vector( 0,-1){100}}
      \put(350,150){\vector( 1, 0){250}}
      \put(600,150){\vector( 0, 1){550}}
      \put(600,700){\vector(-1, 0){300}}
\end{picture}}}
\thinlines
\end{picture}}}
%
%
\end{picture} 
\caption{\sf\small
Branching and weights. Two equivalent generation methods:
(a) with compensation weight for each branch.
(b) with common compensation weight for all branches, 
See text for more explanations how to to transform algorithm (a) into (b).
}
\label{fig:bran-wt2}
\end{figure}
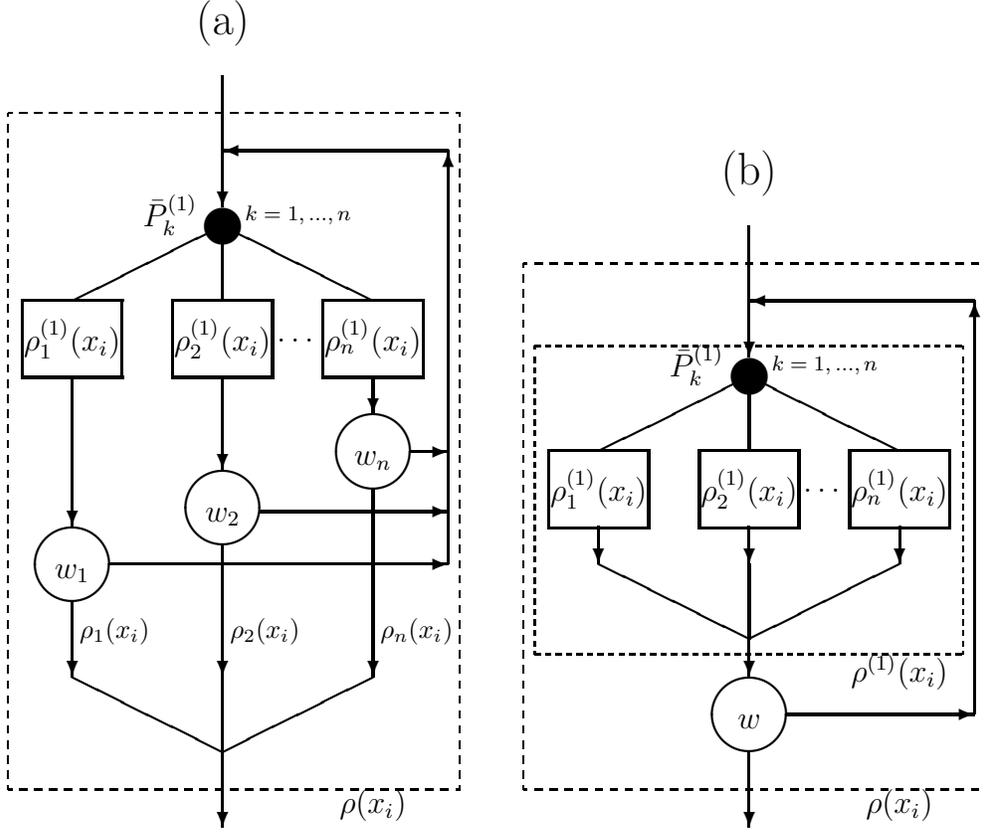

In fig.~\ref{fig:bran-wt2} we transform our algorithm one step further.
In fig.~\ref{fig:bran-wt2}(a) we repeat essentially the algorithm
of fig.~\ref{fig:bran-wt}(b) while in fig.~\ref{fig:bran-wt2}(b)
we have single rejection outside branched part.
The weights $w_k$ and $w$ are related such that two algorithms
are equivalent, that is both algorithms provide the same
distributions and calculate the same integral.
The relations is very simple
\begin{equation}
w
 = \sum\limits_k \bar{p}^{(1)}_k(x)\; w_k(x)
 = \sum\limits_k { \bar{\rho}^{(1)}_k(x) \over \bar{\rho}^{(1)}(x) }\; w_k(x)
 = { \rho(x)  \over  \rho^{(1)}(x) }
 = { \sum\limits_k \rho_k(x)  \over  \sum\limits_k \rho^{(1)}_k(x) }.
\end{equation}
The algorithm
of fig.~\ref{fig:bran-wt2}(b) can be also obtained independently
by combining in a straightforward way the rejection and branching methods.
We proceed as follows: first we simplify
\begin{equation}
\rho(x)  \to  \rho^{(1)}(x)
\end{equation}
and this simplification is compensated by the weight
\begin{equation}
w = { \rho(x)  \over  \rho^{(1)}(x) }
\end{equation}
and for the internal part \fbox{$\rho^{(1)}(x)$} we apply the
branching method as described in section  \ref{sect:branching}.
Consequently, we may apply all standard formulas
for the calculation of the total integral,
for instance $\sigma = \sigma^{(1)} <w>$,
and we do not need
to worry about additional proofs of the correctness
of the algorithm of fig.~\ref{fig:bran-wt2}(b);
we already know that it generates properly the distribution $\rho(x_i)$.

Note that the algorithm of fig.~\ref{fig:bran-wt2}(b) looks more
general than that of fig.~\ref{fig:bran-wt2}(a) in the following sense:
the simplified distribution $\rho^{(1)}$ can be written a sum 
from contributions from all branches
$\rho^{(1)} = \sum\limits_k \rho^{(1)}_k$
and the same is true for $\rho$ in the case (a) while, in general,
it needs not be true in the case (b).
In other words algorithm (a) can be transformed into 
(b) but the transformation in the opposite direction is less obvious.
There is always a trivial transformation of (b) into (a)
in which we set $w_k\equiv w$.
In other words, if in the graph (a) all weights $w_k$ are the same then
we are allowed to contract all rejection loop into a single one
as in graph (b), and vice versa.
This sounds trivial but may be useful in the case of the several
levels of the compensation/rejection weights.

In the case of the variable-weight we simply
omit the rejection return-loops and sum up weights of the events.
Again, since fig.~\ref{fig:bran-wt2}(b) is a direct superposition of
the standard weighting and branching methods all standard rules apply.
It is amusing to observe that in spite of the fact that
the the weights in the two algorithm of figs.~\ref{fig:bran-wt2}
are different, the two algorithm provide exactly the same
distributions and integrals -- only efficiency may differ.
Which one is more convenient or efficient depends 
on the details of a particular problem.
In the next section we shall elaborate on advantages and disadvantages
of the two.

\section{Branching, compensating weights and mapping}

Branching is a very powerful tool in the case of the distribution
with many peaks.
Usually, we are able to split
\begin{equation}
 \rho(x) = \sum\limits_k \rho_k(x)
\end{equation}
in such a way that each $\rho_k(x)$ contains one kind of a spike
in the distribution.
In each branch we generate different spike with help
of the dedicated change of the variables $x_i \to y_i^{(k)}$
such that in
\begin{equation}
 \rho_k(x_i) =  \rho_k(y_i^{(k)})\;
        \left| { \partial x \over \partial y^{(k)} } \right|
\end{equation}
new distribution $\rho_k(y_i^{(k)})$ is completely flat
and the whole spike is located in the Jacobian
function $| \partial y^{(k)} / \partial x  | $.
In the following approximation 
\begin{equation}
\rho_k(x_i) \to \rho^{(1)}_k(x_i) = r^{(1)}_k\;
  \left| { \partial x \over \partial y^{(k)}  } \right|
\end{equation}
the $\rho_k(y_i^{(k)})$ is simply replaced by the constant
residue $r^{(1)}_k$.
The relevant compensating weight reads as follows
\begin{equation}
 w_k = { \rho_k(x_i) \over \rho^{(1)}_k(x_i) }
     = { \rho_k(x_i) \over r^{(1)}_k }\;
       \left| {  \partial y^{(k)} \over \partial x  } \right|
\end{equation}
The approximate cross sections needed for branching
probabilities are
\begin{equation}
\sigma^{(1)}_k 
 = \int\limits_{\Omega} d^n x \rho^{(1)}_k
 = r^{(1)}_k \int\limits_{\Omega_k} d^n y^{(k)}
 = r^{(1)}_k\; V(\Omega_k)
\end{equation}
where $\Omega_k$ is the integration domain 
expressed in the variables $y^{(k)}$ and $V(\Omega_k)$ is simply
the Cartesian volume of the domain $\Omega_k$ in the $y$-space.

Now comes the interesting question:
Which of the two algorithms of fig.~\ref{fig:bran-wt2}
is more convenient and/or efficient.
Finally, the answer will always depend on the individual properties of 
a given distribution $\rho$.
Nevertheless, let us point out some general advantages of the case (a).
In the algorithm of fig.~\ref{fig:bran-wt2}(a) we
need a single weight $w_k$, for the $k$-th branch
from which an event originates.
The distribution $\rho^{(1)}_k$ might be a simple function directly
expressed in terms of $x_i$ but it may happen that the 
Jacobian $|\partial y^{(k)}/ \partial x|$
is a more complicated function which requires the knowledge
of $y^{(k)}_i$ and the whole transformation $y^{(k)}_i \to x_i$.
Of course, it is not the problem for the single branch, as in the case (a),
since in the process of generating an event we calculate primarily 
(generate randomly) the point $y^{(k)}_i$ and we transform it
into $x_i$; the calculation of
this Jacobian is usually a byproduct of the generation process.
The situation in the algorithm of fig.~\ref{fig:bran-wt2}(b)
might be worse because in this case the global weight
\begin{equation}
w = { \rho(x) \over \rho^{(1)}(x) }
= { \sum\limits_k \rho_k(x)  \over \sum\limits_k \rho_k^{(1)}(x) }
= { \sum\limits_k \rho_k(x)  
    \over  
    \sum\limits_k r^{(1)}_k \left|{ \partial y^{(k)} \over \partial x }\right|
  }
\label{wt5b}
\end{equation}
contains in the denominator $\rho_k^{(1)}(x)$
(or Jacobians) for all branches.
Consequently, in some cases we may be forced to perform {\em for each event},
(often quite complicated) transformations 
$x_i \to y_i^{(k)}$ and calculate Jacobians {\em for all branches}.
This is cumbersome, especially if we have large number of branches.
It may also consume a lot of computer time.
Just imagine that due to permutation symmetry we have $N!$
branches -- even if $N$ is some moderately high number the summation
over all branches might consume almost infinite amount of computer time.
The procedure of summation over branches
might be also numerically instable in the case of very strong spikes
in $\rho_k^{(1)}(x)$
because computer arithmetic is usually optimised for 
a given single spike  in a given branch and it might break down
for other branch, unless special labour-hungry methods are employed.

We conclude that the algorithm in fig.~\ref{fig:bran-wt2}(a)
seems to have certain advantages over the algorithm
in fig.~\ref{fig:bran-wt2}(b)
although in many cases the difference might be unimportant
and one might find algorithm (b) more simple 
(it is perhaps easier to explain and document).

\begin{figure}[!ht]
\centering
\htmlimage{scale=1.4}
\setlength{\unitlength}{0.1mm}
\begin{picture}(1400,1200)
%
%
\put(700,1200){\makebox(0,0)[t]{\begin{picture}( 600,1200)
\thicklines
\put(300,1200){\makebox(0,0)[t]{\begin{picture}( 600,200)
      \put(300,200){\vector(0,-1){175}}
      \put(300, 00){\circle*{50}}
      \put(280, 15){\makebox(0,0)[r]{\hbox{ $\bar{P}_k^{(2)}$ }}}
      \put(330, 15){\makebox(0,0)[l]{\hbox{\scriptsize $k=1,...,n$ }}}
\end{picture}}}
\put(300,1000){\makebox(0,0)[t]{\begin{picture}( 600,100)
      \drawline(300,100)(100,0)
      \drawline(300,100)(300,0)
      \drawline(300,100)(500,0)
\end{picture}}}
\put(300,900){\makebox(0,0)[t]{\begin{picture}( 600,100)
      \put(100,100){\makebox(0,0)[t]{\framebox(130,100){$\rho^{(2)}_1(x_i)$}}}
      \put(300,100){\makebox(0,0)[t]{\framebox(130,100){$\rho^{(2)}_2(x_i)$}}}
      \put(500,100){\makebox(0,0)[t]{\framebox(130,100){$\rho^{(2)}_n(x_i)$}}}
      \put(400, 50){\makebox(0,0)[t]{\hbox{ \dots }}}
\end{picture}}}
\put(300,800){\makebox(0,0)[t]{\begin{picture}( 600,300)
      \put(100, 65){\makebox(0,0)[t]{\hbox{ $w_1^{(1)}$ }}}
      \put(100, 50){\circle{100}}
      \put(150, 50){\vector(1,0){450}}
      \put(300,140){\makebox(0,0)[t]{\hbox{ $w_2^{(1)}$ }}}
      \put(300,125){\circle{100}}
      \put(350,120){\vector(1,0){250}}
      \put(500,215){\makebox(0,0)[t]{\hbox{ $w_n^{(1)}$ }}}
      \put(500,200){\circle{100}}
      \put(550,200){\vector(1,0){ 50}}
      \put(600, 50){\vector( 0, 1){550}}
      \put(600,600){\vector(-1, 0){300}}
      \put(100,300){\vector( 0,-1){200}}
      \put(300,300){\vector( 0,-1){125}}
      \put(500,300){\vector( 0,-1){ 50}}
      \drawline(300, 75)(300,0)
      \drawline(500,150)(500,0)
\end{picture}}}
\put(300,500){\makebox(0,0)[t]{\begin{picture}( 600,100)
      \put(100,100){\vector(0,-1){100}}
      \put(300,100){\vector(0,-1){100}}
      \put(500,100){\vector(0,-1){100}}
      \put(110,60){\makebox(0,0)[l]{\hbox{\scriptsize $\rho_1^{(1)}(x_i)$}}}
      \put(310,60){\makebox(0,0)[l]{\hbox{\scriptsize $\rho_2^{(1)}(x_i)$}}}
      \put(510,60){\makebox(0,0)[l]{\hbox{\scriptsize $\rho_n^{(1)}(x_i)$}}}
\end{picture}}}
\put(300,400){\makebox(0,0)[t]{\begin{picture}( 600,100)
      \drawline(100,100)(300,0)
      \drawline(300,100)(300,0)
      \drawline(500,100)(300,0)
\end{picture}}}
\put(300,300){\makebox(0,0)[t]{\begin{picture}( 600,300)
      \put(300,300){\vector(0,-1){100}}
      \put(300,150){\circle{100}}
      \put(300,165){\makebox(0,0)[t]{\hbox{ $w^{(0)}$ }}}
      \put(300,100){\vector( 0,-1){100}} 
      \put(350, 150){\vector( 1, 0){ 300}} 
      \put(650, 150){\vector( 0, 1){1000}} 
      \put(650,1150){\vector(-1, 0){ 350}} 
      \put(310, 60){\makebox(0,0)[l]{\hbox{ $\rho(x_i)$ }}}
\end{picture}}}
%
\put( 15,250){\dashbox{10}(600,875){ }}
\put(500,245){\makebox(0,0)[t]{\hbox{ $\rho^{(1)}(x_i)$ }}}
\thinlines
\end{picture}}}
%
%
\end{picture} 
\caption{\sf\small
Branching and weights. 
Practical example.
}
\label{fig:bran-wt3}
\end{figure}
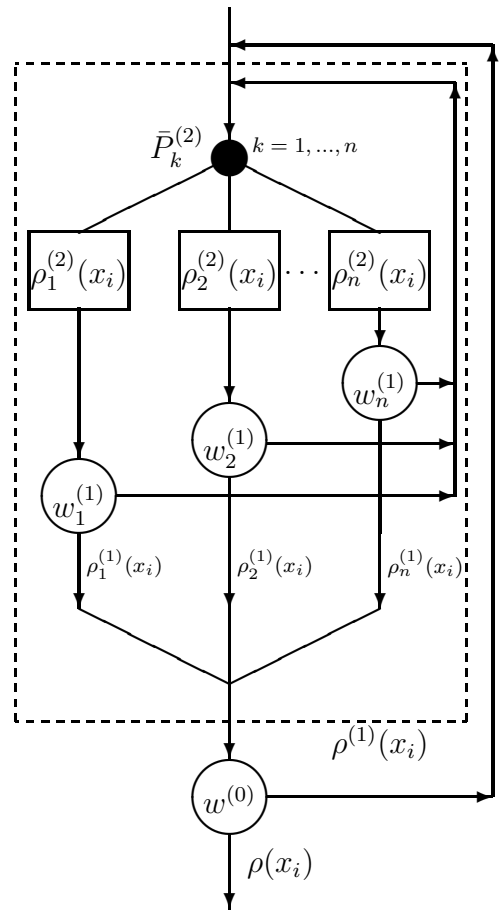

In the two examples of fig.~\ref{fig:bran-wt2} we have
required that $\rho(x)$ can be split {\em immediately} 
into a sum of singular terms, each of them
generated in a separate branch.
In the real life it is often not true
and fig.~\ref{fig:bran-wt3} illustrates more realistic scenario.
Here, before branching can be applied, we make simplification
\begin{equation}
  \rho(x_i) \to  \rho^{(1)}(x_i)
\end{equation}
compensated by the weight
\begin{equation}
  w^{(0)} = {\rho(x_i) \over \rho^{(1)}(x_i) }
          = {\rho(x_i) \over \sum\limits_k \rho^{(1)}_k(x_i) }.
\label{w0}
\end{equation}
This simplification removes fine details from $\rho(x)$
(for example quantum mechanical interferences) which are
numerically unimportant and prevent us from writing $\rho(x)$
as a sum of several positive and relatively simple singular terms.
(Note that in  $w^{(0)}$ we still do not have any Jacobians
and we know nothing about transformation to variables $y_i^{(k)}$!)
The branching is done in the next step for $\rho^{(1)}(x_i)$ and the weights
\begin{equation}
  w^{(1)}_k = {\rho^{(1)}_k(x_i) \over \rho^{(2)}_k(x_i) }.
\end{equation}
compensate for the fact that the Jacobian $|\partial y^{(k)}/ \partial x|$
do not match exactly the $\rho^{(1)}_k(x_i)$, see discussion above.
As in the example of fig.~\ref{fig:bran-wt2}, $w^{(1)}_k$ involves elements
of the calculation for a single branch only (Jacobian!).
The branching probabilities $\bar{P}_k^{(2)}$ are easily calculated using
known integrals $\bar{\sigma}_k^{(2)}$.

The total weight
\begin{equation}
  w = w^{(0)} w^{(1)}_k =
     {\rho(x_i) \over \sum\limits_k \rho^{(1)}_k(x_i) }\;
     {\rho^{(1)}_k(x_i) \over \rho^{(2)}_k(x_i) }
    =
     {\rho(x_i) 
       \over 
      \sum\limits_k \rho^{(1)}_k(x_i) }\;
     {\rho^{(1)}_k(x_i)
       \over
      r^{(1)}_k \left|{ \partial y^{(k)} \over \partial x  }\right| }
\label{local}
\end{equation}
consists of global component  $w^{(0)}$ which knows nothing
about technicalities of generation in the individual branch $k$
and the  local component $w^{(1)}_k$ which bears responsibility for
technicalities of generation in the individual branch $k$
(in particular it may encapsulate cumbersome Jacobian functions).
The lack of sum over $k$ in eq. (\ref{local}) is not a mistake --
the local part of the weight is calculated only for a SINGLE $k$-th branch!!!
This is a great practical advantage 
and such an arrangement of the weights
is generally much more convenient contrary to
the algorithm being the straightforward extension
of the algorithm in fig~\ref{fig:bran-wt2}(b) for which
the weight
\begin{equation}
  w = w^{(0)} w^{(1)} =
     {\rho(x_i) \over \sum\limits_k \rho^{(1)}_k(x_i) }\;
     {\sum\limits_k \rho^{(1)}_k(x_i) \over \sum\limits_k \rho^{(2)}_k(x_i) }
 =
 {\rho(x_i) 
   \over 
  \sum\limits_k \rho^{(1)}_k(x_i) }\;
 {\sum\limits_k \rho^{(1)}_k(x_i)
   \over
  \sum\limits_k r^{(1)}_k \left|{ \partial y^{(k)} \over \partial x }\right| }
\label{global}
\end{equation}
{\em does} include sum all over branches for 
the local part $w^{(1)}$ of the weight.
The efficiency of the two algorithms depends on the details of
the distribution and in order to see which of the above of two
algorithms is more efficient has to be checked case by case.

Another important advantage of the algorithm of eq. (\ref{local})
is that the part of the algorithm generating $\rho^{(1)}_k(x_i)$
can be encapsulated into single subprogram which generated $x_i$
according to $k$-th crude distribution $\rho^{(1)}_k(x_i)$
and provides to the outside world the weight  $w^{(1)}_k$.
The main program does not to need to know more about any details
of the algorithm encapsulated in the subprogram.
The rather annoying feature of the algorithm of eq. (\ref{global})
is that for the construction of the total weight
in the main program we need to know all nuts and bolts of the
sub-generator for $\rho^{(1)}_k(x_i)$,
thus encapsulation cannot be realized,
leading to cumbersome non-modular program.

Finally let us note that the total integral is calculated
with the usual formula
\begin{equation}
  \sigma = \bar{\sigma}^{(2)}\; <w^{(1)} \; w^{(0)}>, \quad
  \bar{\sigma}^{(2)} = \sum\limits_k \bar{\sigma}^{(2)}_k 
\end{equation}
where we understand that for $<w^{(1)}>$ in eq. (\ref{local})
the average is taken over all branches.
For variable-weight events the the weight is $w=w^{(1)} w^{(0)}$
where $w^{(1)}=w^{(1)}_k$ for the {\em actual} $k$-th branch.

\section{Conclusions}
I have described how to combine three elementary Monte Carlo
methods  rejection, branching and change of variables in
the difficult task of generating multi-dimensional distributions.
I have spend some time giving formal mathematical proofs of these
methods, thus providing useful reference for papers describing MC
event generators, where usually authors lack space/time to discuss
such proofs.
I have also discussed in quite some detail advantages and disadvantages
various combinations of branching and rejection methods.
Again, although these aspects may be known to authors of various MC
programs they are practically never discussed.
The most important for practical applications
is probably the discussion on the advantages and disadvantages of the two
arrangements of rejection and branching in fig.~\ref{fig:bran-wt2}.


\end{document}